\documentclass[prb,preprint]{revtex4-1} 

\usepackage{amsmath}  
\usepackage{amsfonts} 
\usepackage{graphicx} 

\usepackage[right]{lineno}
\newcommand {\bfv}[1] {{\boldsymbol {#1}}}

\begin{document}


\title{A simple dynamical model of a freely-falling train of rigid segments}


\author{Tomoaki Itano}
\email{itano@kansai-u.ac.jp}
\affiliation{
  Department of Pure and Applied Physics,
  Faculty of Engineering Science, Kansai University,  Osaka, 564-8680, Japan
}

\author{Masako Sugihara-Seki}
\affiliation{
  Department of Pure and Applied Physics,
  Faculty of Engineering Science, Kansai University,  Osaka, 564-8680, Japan
}


\date{\today}

\date{\today}

\begin{abstract}
In order to elucidate the process underpinning the apparently counterintuitive phenomena observed in the freefall experiments conducted by E. Hamm and J. G\'eminard [Amer. J. Phys. 78, 828 (2010)], we construct a simple dynamical model of a vertically falling train of one-dimensional rigid segments impinging onto an inelastic horizontal plate in three-dimensional space.
Numerically integrating the nonlinear governing equations, we obtain a robust result that the train of rigid segments falls virtually faster than freefall under gravity.
The presented model reproduces the coiling spontaneously formed in the pile, which is considered to be a key mechanism of the phenomenon and is shown to be a consequence of the three-dimensional spiral structure that arises due to dissipative locking in mid-air.
As one of mechanical keys underpinning the apparently counterintuitive phenomena, we will here focus the downward tensile force exerted by the pile.
\end{abstract}

\maketitle 

\section{Introduction}
The freefall of chains is known as an old and solvable but somewhat puzzling exercise involving the variable mass issue, which is often cited in textbooks of classical mechanics\cite{Tho04,Yam57}.
Imagine a chain consisting of a long train of metallic link elements, released initially from a rest state in which it is suspended from one of its ends, falling under gravity onto a pile formed on the ground.
Because of a margin of play between adjacent links, the landing link nearest to the pile is disconnected from the following links of the chain upon colliding with the pile.
The instantaneous impulse of the normal force\cite{Ber98} exerted on the landing link by the pile works only to brake the link itself, and thus compressive stress (pressure) is not transmitted to the successively falling links above it.
Consequently, the landing of the top end of the chain coincides with that of a particle falling freely under gravitational acceleration, as confirmed by experiments.

\begin{figure}[h]
  \centerline{
    \includegraphics[angle=0,height=0.35\textheight]{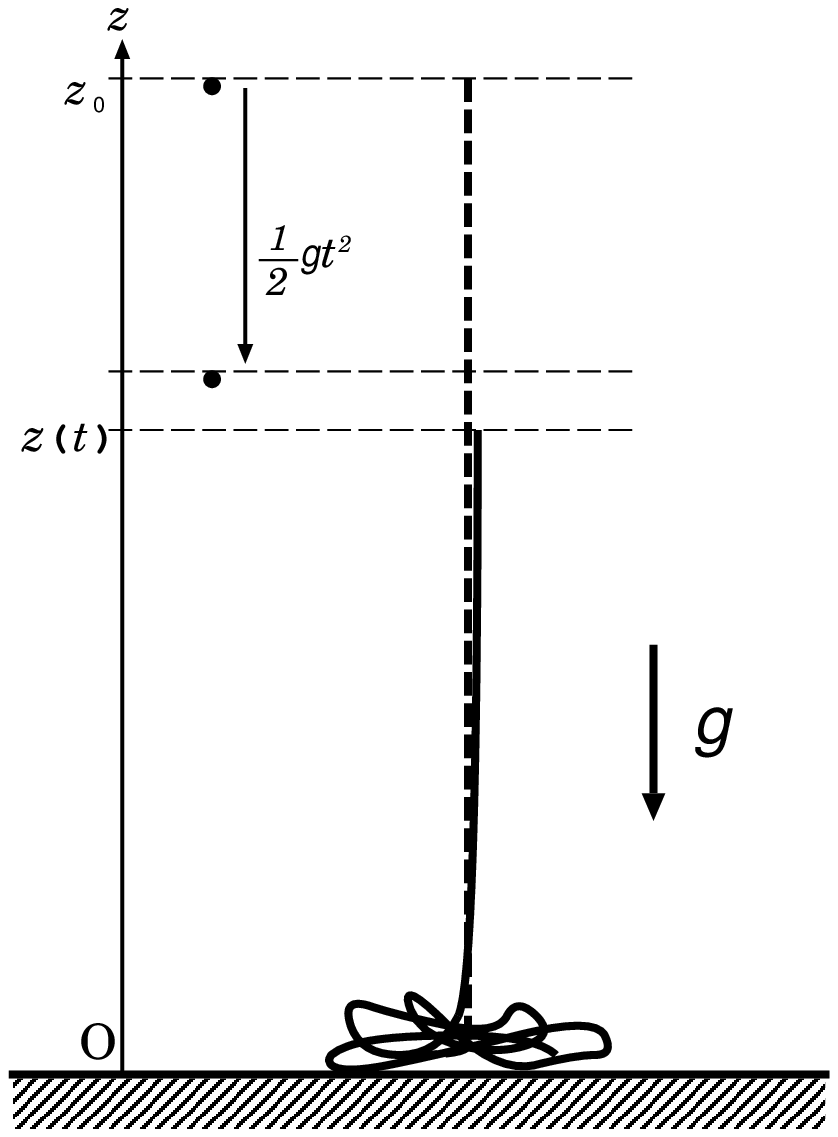}
    \includegraphics[angle=0,height=0.35\textheight]{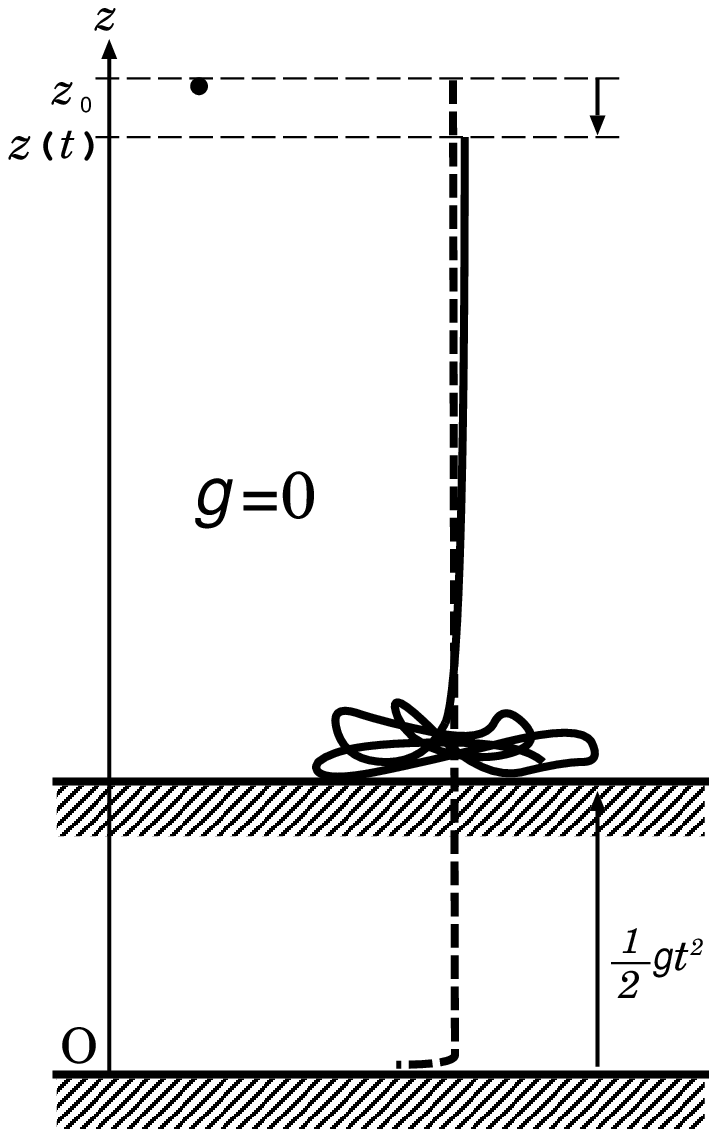}
  }
  \caption{
    (left) A particle and a ballchain fall freely onto an inelastic horizontal plate under gravity.
    (right) The equivalent phenomena. 
    In an inertial frame of reference without gravity, an inelastic plate is accelerated up towards a particle and a ballchain initially at the rest. 
    The ballchain is spontaneously shrunk before the plate collides with the particle.
  }
  \label{fig1}
\end{figure}

Here, we should note that such chains are somewhat of an exception in comparison with other one-dimensional continua.
Due to the steric barrier effect intrinsic to substances, in general, in addition to tension, mechanical pressure will also be internally transmitted in most one-dimensional continua\cite{Ore17}.
This brings about a non-negligible effect on the motion, for example, for wires, ropes, whips\cite{Gor02}, strings, and even threads or fibres, and the understanding of the nature of these one-dimensional continua is of importance in constructions of submarine cables\cite{Air58} or space tethers as well as ground building.
The theory and applications for such a mechanism have been comprehensively presented in Ref.\onlinecite{Ore17}.
With regard to the pressure transmitted internally through one-dimensional continua, our attention was drawn to an apparently counterintuitive phenomena observed in freefall experiments\cite{Ham10} using a particular type of one-dimensional continua called a ``ballchain''\cite{Han12,Big14a,Big14b} or ``Newton's beads'', consisting of metal beads connecting with hinges, which are flexible and dissipative but not so shrinkable or able to wind as much as ordinary chains, because of a smaller margin of play\cite{Gib14}. 
Released from a rest state in which the chain is suspended from one end, a ballchain somewhat unexpectedly completes its falling process in a shorter interval of time than that of a particle simultaneously released at the position from which the ballchain is initially suspended (Fig.\ref{fig1}(left)).
This experimental evidence seems to be apparently paradoxical because one could perform the same experiment even in space without gravity.
The illustration of this result in Fig.\ref{fig1}(right) considers the case when a particle and the top end of the ballchain are initially located at the same height, $z=z_0$, and an inelastic plate is then accelerated upwards. While the particle remains in its initial position, the ballchain is spontaneously shrunk and sucked to the plate somehow via a certain internal tension (net pseudo attractive force between the plate and the ballchain), despite the fact that dissipative collisions can be expected to result primarily in internal compression.

In the present study, we formulate the dynamical system of a vertically falling train of one-dimensional rigid segments impinging onto an inelastic horizontal plate under gravity in three-dimensional space.
Performing a numerical integration of the proposed model's nonlinear governing equations, which to the best of our knowledge has hitherto never been undertaken to the present phenomena, we reproduce the result that the train of rigid segments outruns the ideal freefall.
We found that three-dimensional locking and unlocking events spontaneously occur in the train while it falls through the air, and that the events are transmitted upward through the falling train in the last phase of falling process.
As a mechanical key underpinning the apparently counterintuitive phenomena, we will discuss the downward component of the tensile force exerted on the landing segments pivoting on the pile by the landed segment.

\begin{figure}[h]
  \centerline{
    \includegraphics[angle=0.1,width=0.62\textwidth]{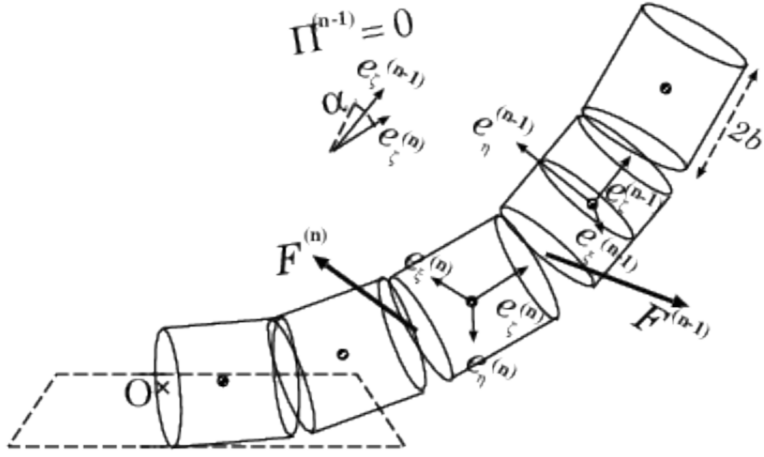}
  }
  \caption{
    The ballchain impinging onto an inelastic horizontal plate, which is modelled as a train of $N$ linked discrete segments, each of which is an unshrinkable rigid body of mass $m$ and length $2b$ with a uniform linear density.
    As illustrated in the top-left corner, the extra force exerted on the $(n-1)$th segment by the $n$th segment, $\bfv{\Pi}^{n-1}(t)$, remains zero whenever $\bfv{e}_\zeta^{(n-1)}(t) \cdot \bfv{e}_\zeta^{(n)}(t) > \cos{\alpha}$, where $\alpha$ is the critical angle.
  }
  \label{fig_model}
\end{figure}

\section{Formulation}
Imagine a ballchain modeled as a train of $N$ linked discrete segments, each of which is an unshrinkable rigid body of mass $m$ and length $2b$ with a uniform linear density.
The principal unit vector and the angular velocity of the $n$th segment ($1\le n \le N$) at time $t$ in the falling process are defined as $\bfv{e}_\zeta^{(n)}(t)$ and $\bfv{\omega}^{(n)}(t)$, respectively.
We assume that initially all the segments are above an horizontal inelastic plate (hereafter termed as the ground), given by the $x-y$ plane of an inertial frame of reference, except with only the bottom end of the $N$th segment of the train being attached to a fixed point on the plane.
We can deduce the equations of momentum and angular momentum of the $n$th segment (see Ref.\onlinecite{Sch97,Tom05} for the motion of rods in two-dimensional space and Ref.\onlinecite{Mah96} for the continuum limit in three-dimensional space) as
\begin{eqnarray*}
  \hspace*{-0.15\textwidth}
  \dot{\bfv{P}}^{(n)} &=& \bfv{F}^{(n)}-\bfv{F}^{(n-1)} + \bfv{\Pi}^{(n)}-\bfv{\Pi}^{(n-1)} -mg\bfv{e}_z\ \ , \\ 
  \dot{\bfv{L}}^{(n)} &=& -b\bfv{e}_{\zeta}^{(n)}\times\bigl( \bfv{F}^{(n)}+ \bfv{F}^{(n-1)} + \bfv{\Pi}^{(n)}+\bfv{\Pi}^{(n-1)}\bigr) \ \ ,
\end{eqnarray*}
where $\bfv{P}^{(n)}=m\dot{\bfv{R}}^{(n)}$ and $\bfv{L}^{(n)}=mb^2 {\cal I}\bfv{\omega}^{(n)}$ are the momentum and the angular momentum of the $n$th segment (${\cal I}$ is the nondimensionalised inertial tensor of a segment), $\bfv{F}^{(n)}$ is the (continuous) interactive force exerting on the $n$th segment by the $(n+1)$th segment at the smooth hinge that connects them.
Additionally, $\bfv{\Pi}^{(n)}$ is the (instantaneous) nondimensionalised extra force exerting on the $n$th segment by the $(n+1)$th segment, which is impulsively generated only at the instant that the hinge between adjacent segments is locked up.
Taking into account that the centre of mass of the $n$th segment, $\bfv{R}^{(n)}(t) = b\bfv{e}_\zeta^{(n)}(t)+2b\sum_{k=n+1}^{N} \bfv{e}_\zeta^{(k)}(t)$, the above governing equations will be converted to a set of evolution equations for $\bfv{e}_\zeta^{(n)}(t)$ and $\bfv{\omega}_\zeta^{(n)}(t)$ ($n=1,2,\cdots,N$) as shown below.

We have demonstrated that a few types of ballchains practically fall faster than freefall in our laboratory, for which typical of specifications are a linear density of 10.2[g/m] and total length of 535 [mm] for 170 beads.
It is remarkable that ballchains are unable to be wound above a curvature specific to their materials, whereas ordinary chains can be wound because of an extensive margin of play that exists between links.
The fact that at least $8 \sim 12$ segments are required to form a circle suggests that adjacent segments of the ballchain can bend along a smooth curve with the angle made between successive links at most up to $\pi/6 \sim \pi/4$. 

Reflecting the aforementioned fact, we thus assume for simplicity that the extra force, $\bfv{\Pi}^{(n)}$, is expressed as the gradient of a box-style infinite potential well, that is, the magnitude of the force is zero whenever the angle of $\bfv{e}_{\zeta}^{(n)}$ and $\bfv{e}_{\zeta}^{(n-1)}$ is within a critical angle, $\alpha$, while if this angle reaches $\alpha$ the two segments are instantaneously locked together by inelastic collision at the hinge so as to dissipate their kinetic energy with respect to their relative rotations.
Thus, the angular velocity of two segments, $\bfv{\omega}_\zeta^{(n)}$ and $\bfv{\omega}_\zeta^{(n-1)}$, becomes equal after the collision, and is thereafter kept to be identical unless some anti torque produced by adjacent segments unlocks them.
An alternative way of introducing certain energy dissipation rate with regard to the relative rotation of two adjacent segments was previously proposed in Ref.\onlinecite{Sch97}.
We would be able to obtain analogous results in the present study.
The scope of Ref.\onlinecite{Sch97} was, however, restricted to motion of ballchains in two-dimensional space, unlike the present three-dimensional investigation.

Moreover, in order to model the inelastic collisions of the bottom segment against the ground, we account that an impulsive extra force exerted on the $N$th segment, $\bfv{\Pi}^{(N)}$, is generated at the same time as the upper end of the $N$th segment reaches the $x-y$ plane in the falling process, i.e., at the instant that $\bfv{e}_\zeta^{(N)}(t) \cdot \bfv{e}_z =0$.
The rest of the $N$th segment is led by this inelastic collision against the ground, which yields a reduction in the numerical cost of the calculations, by decreasing the number of segments of the train to be calculated, which is equivalent to the conversion $N \to N-1$ in the numerical scheme.
Algebraically eliminating $\bfv{F}^{(n)}$ from the above equations as long as $\bfv{\Pi}^{(n)}=0$, we obtain the following set of Euler's rotation equations for the rigid segments,
%
%
\begin{eqnarray*}
  \frac{d}{dt}\Bigl({\cal I}\bfv{\omega}^{(n)}\Bigr)  &=&  \bfv{Q}^{(n)}  -\frac{(2n-1)g}{b}\bfv{e}_{\zeta}^{(n)} \times \bfv{e}_z \ \ \ ,  
\end{eqnarray*}
where $\bfv{Q}^{(n)}$ is a family of nonlinear functions of $\bfv{e}_{\zeta}^{(k)},\bfv{\omega}^{(k)},\dot{\bfv{\omega}}^{(k)}$ for $k=1,\cdots,N$.
With the aid of complementary equations, $\dot{\bfv{e}}_{\zeta}^{(n)}=\bfv{\omega}^{(n)}\times \bfv{e}_{\zeta}^{(n)}$, the motion of the falling ballchain can be solved by numerically integrating the set of ordinary differential equations.
When a collision between adjacent segments or a collision between the bottom segment and the ground occurs, the corresponding impulsive extra force affects the ballchain motion through the time integration of equation of motion within an infinitesimal interval.
The system can thus be perceived as a transient nonlinear dynamical system with dissipation.

\begin{figure}[h]
  \centerline{
    \includegraphics[angle=0,width=0.75\textwidth]{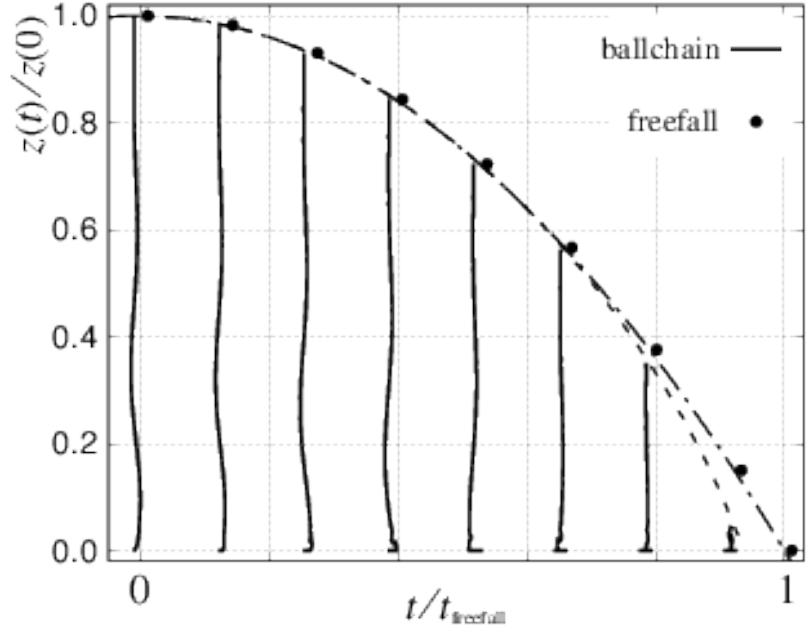}
  }
  \caption{
    Stroboscopic pictures of a falling ballchain taken at a constant time interval, $0.13 t_{\rm freefall}$, with the dashed curve showing the time series of the $z$ coordinate of the top end of the falling ballchain given at $(N,\alpha)=(128,\frac{\pi}{4})$.
    For reference, a particle freely falling from $z(0)$ at $t=0$ is indicated $\bullet$ (interpolated by a dash-dotted curve), which collides the floor at $t/t_{\rm freefall}=1$, where $t_{\rm freefall}=\sqrt{2z(0)/g}$.
    The falling process of the ballchain takes $t=0.94 t_{\rm freefall}$, which is in accordance with the experimental result reported in Ref.\onlinecite{Ham10} and trial demonstrations in our laboratory.
  }
  \label{fig2}
\end{figure}

\section{Results}
Introducing several types of three-dimensional smooth perturbation into the orientation distribution of the segments initially at the rest, the numerical integration of the equation of motion has been performed under given parameters.
Note that the dynamics of the system is determined only by the two nondimensional parameters, $(N,\alpha)$, while the other parameters, $(m,b,g)$, to scale the magnitudes of mass, length, time, have no essential influence on the dynamics.
The initial height of the top end of the ballchain, $z(0)$, is slightly less than the total length of the ballchain, $2bN$, due to the slackness introduced at the initial allocation, $z(0) \lesssim 2bN$.

Fig.\ref{fig2} is a typical sequence of snapshots taken at constant time intervals of a falling ballchain numerically integrated for $(N,\alpha)=(128,\pi/4)$, which elucidates that the ballchain completes the falling process earlier than that of freefall. 
The abscissa is the time elapsed after the release nondimensionalised by $t_{\rm freefall}$, which is the interval for a freely falling particle to travel the distance $z(0)$.
The dashed curve in the figure is the time series of the $z$ coordinate of the top end of the falling ballchain.
It can be seen from the figure that the difference between the top end of the ballchain and the freely-falling particle, which is not remarkable at the initial stage of the fall, increases abruptly in the last third of the process, $0.6 \lesssim t/t_{\rm freefall}$.
Extrapolating the curve from the terminal stage of the numerical result, we estimate that the ballchain completes its falling at $t/t_{\rm freefall} \approx 0.94$ for the cases shown, which provides an excellent agreement with the available experimental evidence reported in Ref.\onlinecite{Ham10}.

\begin{figure}[h]
  \centerline{
    \includegraphics[angle=0,width=0.68\textwidth]{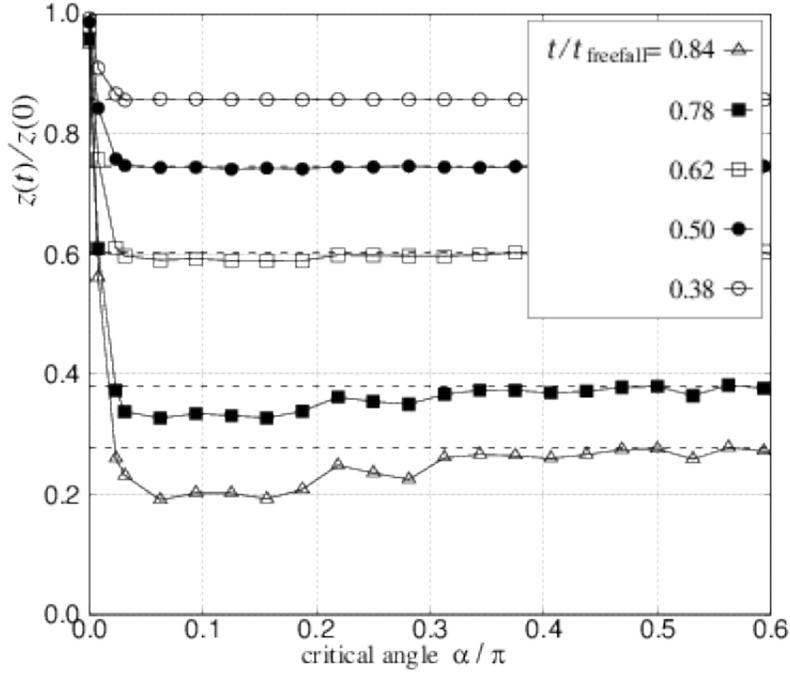}
  }
  \caption{
    The height of the upper end of the ballchain at the several times in $0.38 \le t/t_{\rm freefall} \le 0.84$. The numerical integrations were performed under the same initial condition but using different values of $\alpha$.
  }
  \label{fig5}
\end{figure}

According to numerical investigations of the present minimal model, the fact that ballchains fall faster than freely-falling particle is found to be a robust result for a wide range of parameters despite the results' dependence on initial conditions.
We should emphasise that the simulations also showed that the results are specially sensitive to the critical angle, $\alpha$; the acceleration can be equal to or less than the freefall in case that the given $\alpha$ is too small or large compared to $\pi/6$.
This implies the existence of an optimal value in $\alpha$.
Fig.\ref{fig5} shows the height of the upper end of the ballchain at the representative instances in $0.38 \le t/t_{\rm freefall} \le 0.84$, obtained from numerical integration under the same initial condition ($N=256$) but using different values of $\alpha$.
The figure indicates a non-negligible dependence on $\alpha$ of the height of the top-end $z(t)$ at a fixed time $t$.
This implies that the optimal value exists between $0 < \alpha \le \pi/3$, rather than the case of $\alpha=\pi/4$ illustrated in Fig.\ref{fig2}, which will be investigated in future work.
At a larger $\alpha$ ($\alpha>\pi/2)$, the top segment falls down almost as fast as a freely-falling particle as shown in the figure, and so this case can be considered to correspond to that of a regular chain.
For regular chains, the landing process of the bottom segment on the pile does not strongly affect that of the following falling segment.
On the other hand, the limit $\alpha=0$ corresponds to the case in which all the segments are totally locked and the chain falls down as an axed tree, which requires an infinite time to complete when there is no perturbation to the initial allocation.
From the dynamical point of view, the existence of an optimal $\alpha$ suggests that the acceleration of the ballchain can be attributed to the correlation of directions of adjacent segments landing on the ground, where extra force $\bfv{\Pi}^{(n)}$ locally generates (also see Fig.\ref{fig3}).
In addition, note that the difference from the freely falling particle becomes remarkable at the latter part of the falling process after $t / t_{\rm freefall} \sim 0.62 $ for the optimal $\alpha$.


\begin{figure}[h]
  \centerline{
    \includegraphics[angle=0,width=0.75\textwidth]{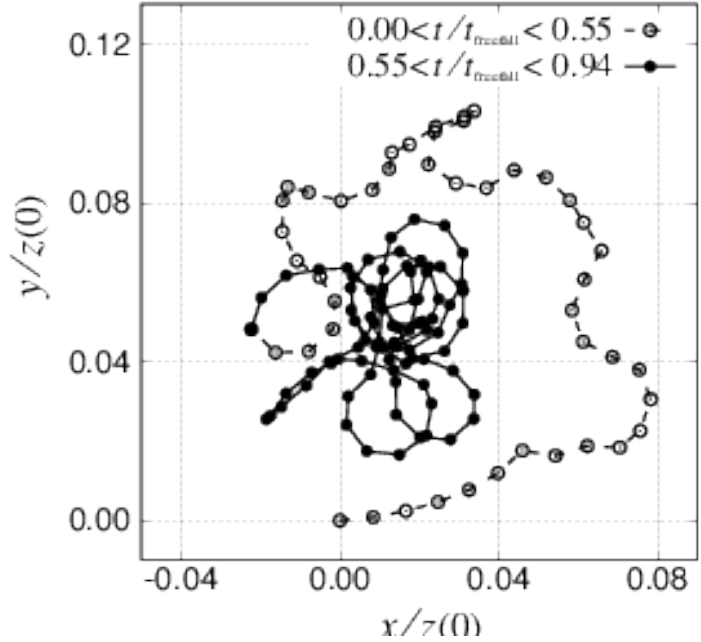}
  }
  \caption{
    The top view of the pile of the ballchain formed on the ground.
    The parameter and the initial allocation are the same as for Fig.\ref{fig2}.
    The coiling of the ballchain is striking especially in the latter interval of the falling process, when it is significantly accelerated in comparison with freefall.
  }
  \label{fig4}
\end{figure}

In dissipative collisions of one-dimensional continua onto an inelastic horizontal plane, both coiling and folding patterns have been commonly observed\cite{Mah96,Rib06}.
However, it is not self-evident why these systems of one-dimensional continua prefer particular coiling and folding patterns.
The present dynamical model numerically reproduces the phenomenon that ballchains spontaneously form either a three-dimensional spiral or twisted shape in the air above the pile, which is required for the coiling that then occurs on the pile.
The formation of the spiral in the air is actually captured as a tiny wiggling seen around the $z(t)/z(0)\sim 0.05$ in Fig.\ref{fig2}.
An example of the top view of the pile formed on the ground is shown in Fig.\ref{fig4}, where landed segments are distinguished before and after $t/t_{\rm freefall}=0.55$, the time beyond which the falling acceleration significantly overtakes that of freefall.
While the orbit of the pile formed in the former interval is meandering, it is striking that the orbit then coils up in the latter interval.
It is observed that, in some cases depending on the initial condition, the orbits are swung from side to side like a figure of eight.
In the present study, the coiling formed in the pile is considered as a byproduct of the three-dimensional spiral structure in the train of segments due to dissipative locking on mid-air, which may be an attractor of this transient dissipative dynamical system.

\section{Discussion}
The ``conversion'' of internal compressive (attributed to the {\it upward} normal reaction force exerted by the ground) to tensile stresses proposed by Grewal et al.\cite{Gre11} could be considered as one possible explanation of falling ballchains.
According these authors, the angular velocity of the bottom segment increased by the upward normal force  results in sucking downward the following segments as described in Fig.2.13 in p.69 in Ref.\onlinecite{Ore17}.
They demonstrated experimentally that a handmade toy rope-ladder consisting of cylindrical rods linked between a couple of hanging threads, instead of a ballchain, falls faster than freefall under gravity as if it is virtually sucked by the ground.
Each rod is tied to the threads at its ends with an angle slightly inclined to the horizontal, so that the collisions of the rods to the ground produce internal tension through the threads resulting in the acceleration of falling rope-ladder.
However, we should also note the fact that the rope-ladder consisting of the double threads are intrinsically unable to transmit internal pressure upwards after all, which is structurally distinct from ballchains.
That is, internal pressure attributed to the normal force exerted by the ground
, which could be a source of deceleration of falling ballchain, is not inherent in threads of the rope-ladder.
Thus, it is not obvious whether the proposed acceleration mechanism for falling rope-ladders is directly applicable to that of ballchains.

\begin{figure}[h]
  \centerline{
    \includegraphics[angle=0,width=0.60\textwidth]{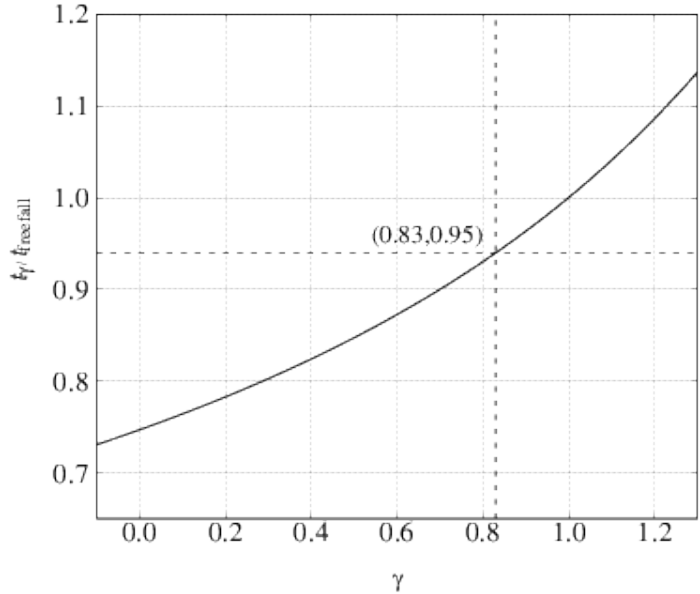}
  }
  \caption{
    The relation of the nondimensionalised time interval to complete the falling process  against a model parameter $\gamma$ in Ref.\onlinecite{Ham10}. The model provides a prediction of  $\displaystyle \frac{t_{\gamma}}{t_{\rm freefall}}>0.74$, which is satisfied by both the present numerical results and the experimentally obtained results.
  }
  \label{fig_Beta}
\end{figure}

According to Hamm and G\'eminard\cite{Ham10}, the vertical component of momentum equation and the horizontal component of angular momentum equation may account for the apparently counterintuitive phenomena.
Estimating the order of the time variation of angular momentum under the assumption that the falling ballchain coils up on the pile scaled with a representative length, they deduced that the {\it upward} normal force exerted on the ballchain via the bottom segment on the pile by the ground is approximated as $\gamma \mu \dot{z}^2$.
Here, $\mu$ is linear mass density, and, a geometrical factor $\gamma$ is given by the shape of the curl.
The magnitude of $\gamma$ corresponds to the ratio of the normal reaction force exerted by the ground to brake the landing bottom segment ($N_{\rm react}$) to the loss of the momentum of the landing bottom segment ($N_{\rm all}$), that is, $\gamma=N_{\rm react}/N_{\rm all}$ (for freefall $\gamma=1$). 
Note that the $1-\gamma$ portion of $N_{\rm all}$ (the loss of the momentum of the landing bottom segment) is consumed to the acceleration of the falling train except for the landing bottom segment, due to the internal tension of the segments on the pile.
Therefore, $\gamma=0$ corresponds to a variable mass case in that the landing bottom segment is braked only through the vertically upward force exerted by the falling train except for the landing bottom segment.
Analytically integrating the differential equation proposed in Ref.\onlinecite{Ham10}, we obtained the interval required to complete the falling for an arbitrary $\gamma$ as
\begin{equation*}
  \frac{t_{\gamma}}{t_{\rm freefall}} = \frac{1}{2\sqrt{3-2\gamma}} B \Bigl(\frac{2-\gamma}{3-2\gamma},\frac{1}{2}\Bigr) \ \ ,
\end{equation*}
where $B$ is the conventional beta function. 
Fig.\ref{fig_Beta} shows the nondimensional interval $\displaystyle \frac{t_{\gamma}}{t_{\rm freefall}}$ against $\gamma$.
Although they concluded that $\gamma$ is 0.83 in their experimental results, the magnitude of $\gamma$ could fit an arbitrary experimental value within $0.74 \le t_{\gamma}/t_{\rm freefall} \le 1$.

\begin{figure}[h]
  \centerline{
    \includegraphics[angle=0,width=0.65\textwidth]{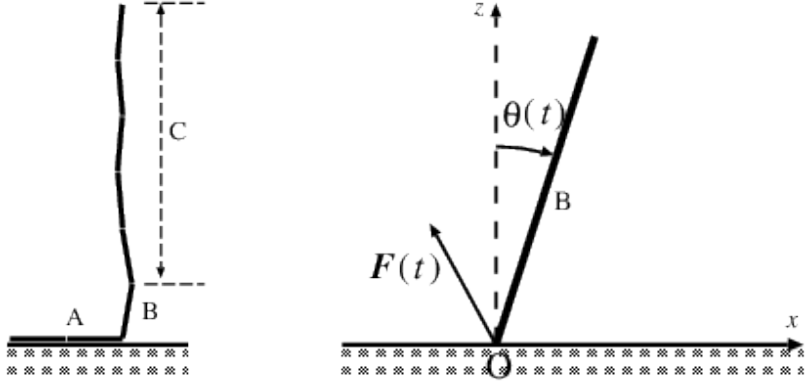}
  }
  \caption{
    (left) The falling ballchain consists of the three parts, ``A'' the landed segments which have already been laid on the ground works as an anchor, ``B'' the landing segment pivoting on the end that attaches to A, ``C'' the falling segments in the air. (right) A linear rigid body falls down under the gravity and the reaction force exerted at the origin on the ground.
  }
  \label{fig_mech}
\end{figure}

We will here review from a mechanical point of view, the motion of a linear rigid body falling down with its bottom end attached to the origin on the ground (see Fig.\ref{fig_mech}(right)).
This is because such a mechanism may fairly represent the motion of the landing segment ("B" in Fig.\ref{fig_mech}(left)) pivoting at the bottom of the falling ballchain, if it is not largely affected by the train of segments falling in the air ("C" in Fig.\ref{fig_mech}(left)) connected from its top end and if the segment which has been already landed on the ground ("A" in Fig.\ref{fig_mech}(left)) does not move as an anchor on the ground due to its own weight, so that the anchor segment "A" fixes the pivoting position of "B" at the origin, O.
The azimuthal angle $\theta$ of the principal direction of the landing segment B increases with time $t$.
Assuming that the length and the mass of a segment B is $2b$ and $m$, respectively, under the magnitude of gravitational acceleration $g$, we can analytically solve $\theta(t)$ following the textbooks\cite{Tho04,Yam57}, and the normal component of the reaction force exerted to B is given as a function of $\theta$ by
\begin{equation*}
  \bfv{F}(t)\cdot \bfv{e}_z =
  \frac{mg}{4}\Bigl\{ \bigl(1-3\cos{\theta}\bigr)^2 -6\bigl( 1- \cos{\theta_0}\bigr) \cos{\theta} - 4s \cos{\theta} \Bigr\},
\end{equation*}
where $\theta_0=\theta(0)$ and $\displaystyle s=\frac{b}{g} |\dot{\theta}(0)|^2$.
The impulse exerted on the landing segment B by the landed anchor A within the falling interval $\theta(0)<\theta<\pi/2$, $\Delta p(\theta_0,s):=\displaystyle \int_{\theta_0}^{\pi/2} \dot{\theta}^{-1} \bigl( \bfv{F}(t)\cdot \bfv{e}_z\bigr) d\theta$, can be numerically calculated.
Although the value of the impulse can be positive or negative, in general, depending on the parameters $(\theta_0, s)$, it is negative over a sufficiently large value of $s$ at any $\theta_0$.
Thus, the landed anchor starts to pull downwards the falling ballchain when the falling speed is over a critical value after some acceleration, as if it is sucked to the ground; the landed segments can practically pull downward the segments falling in the air, even if the conversion mechanism due to the upward reaction force proposed in Ref.\onlinecite{Gre11} is not realized.

For simplicity, suppose that the falling speed of the top end of the ballchain, $|\dot{z}(t)| \approx g t$ and it also equals to $2b |\dot{\theta}(0)|$, which is the circumference speed of the top end landing segment B, then $\frac{1}{2}g t^2 = 2b s$.
Note that the value of $s$ corresponds to the number of the landed segments.
We obtained that initially $\Delta p$ could be positive, but then changes to negative and eventually ($t/t_{\rm freefall}>0.5$) becomes comparable to the loss of the momentum of falling segments $C$ within the interval in $s \to s+1$.
Therefore, the ``sucking to the ground'' may work effectively if the coiling on the pile is kept to be continuous, where the value of $\dot{\theta}(0)$ for each segment, which successively work as pivoting segments, increases as time elapses.

\begin{figure}[h]
  \centerline{
    \includegraphics[angle=0,width=0.70\textwidth]{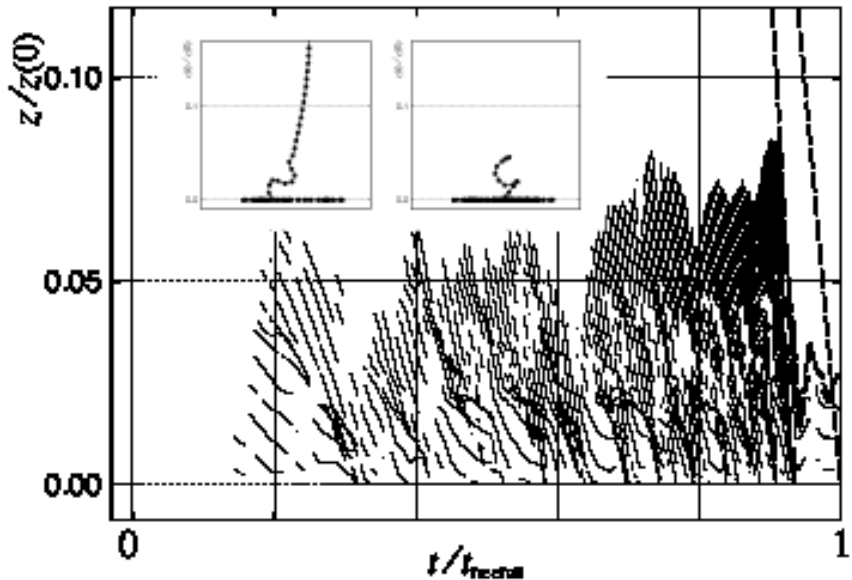}
  }
  \caption{
    The time series of the height of locked adjacent segments.
    Assemblies of locked adjacent segments emerging within regions indicated by the oblique lines are intermittently unlocked within blank speckles appearing inside a region.
    The lower dashed curve at $t>0.94$ corresponds to the $z$ coordinate of the top end of the falling ballchain, which is identical to that in Fig.\ref{fig2}.
    Insets show the side views of ballchain taken at $t/t_{\rm freefall}=0.66$ and $0.92$, parts of which are locked and autonomously form ``inclined rods of a rope-ladder'' proposed by Ref.\onlinecite{Gre11}.
  }
  \label{fig3}
\end{figure}


The principal vectors of the segments, which are initially aligned close to the vertical direction, end up laid out horizontally under the falling process.
Generated at the impingement to the pile, the reaction force $\bfv{F}^{(N)}$ exerted on the landing bottom segment by the ground changes its orientation from the vertical to the horizontal while braking it to rest.
In the present numerical scheme, a couple of adjacent segments of a ballchain, not permitted to bend over a given critical angle, are dissipatively locked together as a result of an inelastic rotating collision at the hinge.
Thus, the impulsive forces generated at the impingement to the ground could be transmitted upward through the successively falling segments following the bottom segment to the top end.
On the other hand, the accelerated falling speed eventually defies the upward transmission speed of the torque produced by the locking.
Such a locking of segments is actually captured as a tiny zigzag seen at the bottom of the falling ballchain in insets of Fig.\ref{fig2}.
Here, we recorded the locked adjacent segments and plotted the time series of their coordinates in Fig.\ref{fig3}, where the centres of mass of locked segments are plotted as a dot.
Note blank speckle regions under the lower dashed curve in the figure corresponds to where segments fall freely  almost independently, neither compressed nor wound by the adjacent segments, without locking with adjacent segments.
As time elapses, several adjacent independent segments above the pile develop into a group of segments, while a group of segments divided to a few groups intermittently (blank speckles appears inside a region indicated by oblique lines).
The intermittently repeated locking and unlocking by an assembly of segments may be regarded as the autonomous formation of ``inclined rods of a rope-ladder'' proposed by Ref.\onlinecite{Gre11} while the ballchain is falling through the air.
However, these groups of locked and unlocked segments are formed only at the bottom and never develops upward $z/z(0)>0.1$.
This fact is relevant to the experimental observation reported by Hamm and G\'eminard \cite{Ham10}; thus we do not find the deformation of the chain at the bottom excites transverse waves that eventually climb upwards along the falling chain.

\section{Summary}
The motion of ballchains falling vertically and impinging onto a horizontal inelastic plane was formulated as a set of nonlinear ordinary differential equations, which were numerically solved.
The counterintuitive experimental result that ballchains fall faster than freefall under gravity was reproduced by the present numerical scheme.
From trials of integration under several different initial conditions, it was confirmed that the evidence is robust, but that it is especially sensitive to the critical angle, $\alpha$, at which a pair of adjacent segments of the ballchain collide and dissipate their kinetic energy with respect to their relative rotations.

We discussed why acceleration is possible, even if the conversion mechanism due to the upward reaction force proposed in Ref.\onlinecite{Gre11} is not realized.
When the falling speed is over a critical value, it is possible that the landed segment ``anchor'' in practice pulls down the falling ballchain.
More precisely, rather than the ballchain being sucked to the ground, it can be said that the landed segments of ballchain suck downward the train of segments falling in the air.

It was also found that a  three-dimensional locking and unlocking of adjacent assembly of segments occur intermittently just above the pile, while the ballchain is falling through the air.
The locking and unlocking events originate in the torque generated at the impingement incidence of the bottom of the falling segment to the pile, with this torque transmitted upwards against the falling.
The falling speed is accelerated to eventually defy the upward transmission speed of the torque produced by the locking.
The coiling formed on the pile is considered as a byproduct of the three-dimensional spiral structure in the train of segments due to dissipative locking in mid-air.

The falling ballchain has turned out to be a surprisingly rich phenomenon involving nonlinear dissipative dynamics, although it can be formulated and explained  by classical mechanics.

\acknowledgments 
We thank Dr Yokoyama, Prof. J. Seki and Prof. N. Sugimoto for comments on an earlier draft of the manuscript, and Dr D.P. Wall for improving the manuscript.
T.I. is grateful for the financial support in part by the Kansai University Subsidy for Supporting Young Scholars, 2017.





\bibliographystyle{apsrev4-1}
\bibliography{blc01}

\end{document}